\documentclass[twocolumn,showpacs,preprintnumbers,superscriptaddress,amsmath,amssymb,prl]{revtex4}
\usepackage{graphicx}% Include figure files
\usepackage{dcolumn}% Align table columns on decimal point
\usepackage{bm}% bold math
\usepackage{quantum}
\usepackage{amsmath}
\begin{document}

\preprint{APS/123-QED}

\title{High Efficiency Photon Number Detection for Quantum Information Processing}
\author{Edo Waks}
\affiliation{Quantum Entanglement Project, ICORP, JST, E.L.
Ginzton Laboratories, Stanford University, Stanford, CA 94305 }
\author{Kyo Inoue}
\affiliation{Quantum Entanglement Project, ICORP, JST, E.L.
Ginzton Laboratories, Stanford University, Stanford, CA 94305 }
\affiliation{NTT basic research, Atsugi, Kanagawa, Japan}
\author{William D. Oliver}
\affiliation{Quantum Entanglement Project, ICORP, JST, E.L.
Ginzton Laboratories, Stanford University, Stanford, CA 94305 }
\author{Eleni Diamanti}
\affiliation{Quantum Entanglement Project, ICORP, JST, E.L.
Ginzton Laboratories, Stanford University, Stanford, CA 94305 }
\author{Yoshihisa Yamamoto}
\affiliation{Quantum Entanglement Project, ICORP, JST, E.L.
Ginzton Laboratories, Stanford University, Stanford, CA 94305 }
\affiliation{NTT basic research, Atsugi, Kanagawa, Japan}

\date{\today}% It is always \today, today,

\begin{abstract}

The Visible Light Photon Counter (VLPC) features high quantum
efficiency and low pulse height dispersion. These properties make
it ideal for efficient photon number state detection.  The ability
to perform efficient photon number state detection is important in
many quantum information processing applications, including recent
proposals for performing quantum computation with linear optical
elements.  In this paper we investigate the unique capabilities of
the VLPC.  The efficiency of the detector and cryogenic system is
measured at 543nm wavelengths to be 85$\%$.  A picosecond pulsed
laser is then used to excite the detector with pulses having
average photon numbers ranging from 3-5.  The output of the VLPC
is used to discriminate photon numbers in a pulse.  The error
probability for number state discrimination is an increasing
function of the number of photons, due to buildup of
multiplication noise. This puts an ultimate limit on the ability
of the VLPC to do number state detection.  For many applications,
it is sufficient to discriminate between 1 and more than one
detected photon.  The VLPC can do this with 99$\%$ accuracy.

\end{abstract}

\maketitle

\section{Introduction}
Optical quantum information processing is one of the most rapidly
developing segments of quantum information to date. The photon
offers many distinct advantages over other implementations of a
quantum bit (qubit).  It is very robust to environmental noise,
and can be transmitted over very long distances using optical
fibers.  For this reason the photonic qubit is the exclusive
information carrier for quantum cryptography applications.  Recent
theoretical developments have also shown that single photons,
combined with only linear optical components and photon counters,
can be used to implement scalable quantum computers.

At the heart of any optical quantum information processing
application is the ability to detect photons.  Photon counters are
an essential tool for virtually all quantum optics experiments.  A
photon counter absorbs a single photon, and outputs a macroscopic
current that can be processed by subsequent digital circuits.   To
date, photomultiplier tubes (PMTs) and avalanche photodiodes
(APDs) are the most common photon counters.  In a PMT, a photon
scatters a single electron from a photocathode.  The electron is
multiplied by successive scattering off of dynodes in order to
generate a macroscopic current.  PMTs are known to have superb
time resolution and low pulse height dispersion, yet they
typically suffer from low detection efficiencies.  Optimal quantum
efficiencies for a PMT typically do not exceed 40$\%$.  Avalanche
photodiodes feature higher quantum efficiencies.  In an APD, a
photon creates a single electron hole pair in a semiconductor pn
junction.  An avalanche breakdown mechanism multiplies this
electron-hole pair into a large current.  APDs can have quantum
efficiencies as high as 75$\%$.  The main limitations of APDs is
that they have a relatively long dead time (~35ns), and large
pulse height dispersion.  If two photons are simultaneously
absorbed by the APD, the output pulse will not differ from the
case when only one is absorbed.  Thus, APDs cannot distinguish
between one and more than one photon if all of the photons land
within the dead time of the detector.  We refer to such detectors
as threshold detectors.

Recently, a new type of photon detector, the Visible Light Photon
Counter (VLPC), has been shown to have some unique capabilities
that conventional PMTs and APDs don't have. The VLPC features high
quantum efficiencies (~94$\%$), and low pulse height
dispersion~\cite{KimYamamoto97,TakeuchiKim99}.  This latter
property makes the VLPC useful for photon number
detection~\cite{KimTakeuchi99}. Unlike an APD, if two photons are
simultaneously absorbed by the VLPC, the detector outputs a
voltage pulse which is twice as high. This behavior continues for
higher photon numbers. Thus, the voltage pulse of the VLPC carries
information about photon number. We refer to this type of detector
as a photon number detector, in contrast to threshold detectors
discussed previously.

Photon number detection is very useful for quantum information
processing.  It has applications in quantum cryptography,
particularly in conjunction with parametric down-conversion.  It
is also an important element for linear optical quantum
computation (LOQC) as proposed by Knill, Laflamme, and
Milburn~\cite{KnillLaflamme01}. Many of the basic building blocks
for this proposal fundamentally rely on the ability to distinguish
between one and more than one photon with high quantum
efficiency~\cite{BartlettDiamanti02}.

There are several unique aspects of the VLPC which allow it to do
photon number detection. First, the VLPC is a large area detector,
whose active area is about 1mm in diameter. When a photon is
detected, a dead spot of several microns in diameter is formed on
the detector surface, leaving the rest of the detector available
for subsequent detection events. If more than one photon is
incident on the detector, it will be able to detect all the
photons as long as the probability that multiple photons land on
the same location is small. This is a good approximation if the
light is not too tightly focussed on the detector surface.  In
this respect the VLPC is similar to a large array of beamsplitters
and threshold detectors.

There is, however, one major distinction between the VLPC and a
large detector array. In an array, we can address the signal from
each counter individually.  In contrast, we cannot individually
access each spot on the VLPC surface.  Instead, the current from
the entire detector is summed and accessed through a single
output. We must use the height or area of the output pulse to
infer the photon number. This makes the noise properties of the
detector critical for photon number detection. When independent
noisy voltages are summed the noise builds up.  This degrades the
number resolution capability of the detector, and ultimately puts
a limit on the number of incident photons that can be resolved. It
is important to measure this limitation in order to assess the
capability of the detector to perform quantum information
processing tasks.  For this we need to consider the noise
properties of the VLPC.

The noise properties of a photon counting system are ultimately
limited by the internal multiplication noise of the detector.
Photon counters typically rely on an internal multiplication gain
to create a large current spike from a single photoionization
event. These gain mechanisms create internal multiplication noise,
meaning that the current spikes generated by a photodetection
event will fluctuate in height and area.  In order to do accurate
photon number detection, this multiplication noise must be low.
Fortunately, the VLPC has been measured to have nearly noise-free
multiplication~\cite{KimYamamoto97}.

\section{VLPC operation principle}

  \begin{figure}
    \centerline{\scalebox{0.4}{\includegraphics{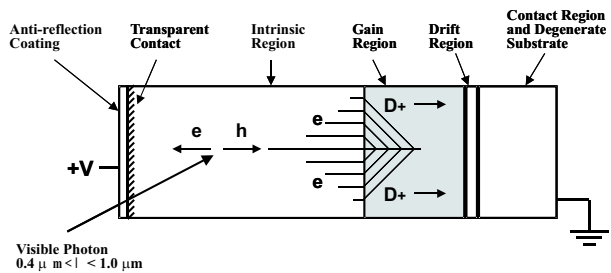}}}
    \caption{Schematic of structure of the VLPC detector}\label{fig:VLPCschematic}
  \end{figure}

Figure~\ref{fig:VLPCschematic} shows the structure of the VLPC
detector.  Photons are presumed to come in from the left.  The
VLPC has two main layers, an intrinsic silicon layer and a lightly
doped arsenic gain layer.  The top of the intrinsic silicon layer
is covered by a transparent electrical contact and an
anti-reflection coating. The bottom of the detector is a heavily
doped arsenic contact layer, which is used as a second electrical
contact.

A single photon in the visible wavelengths can be absorbed either
in the intrinsic silicon region or in the doped gain region. This
absorption event creates a single electron-hole pair. Due to a
small bias voltage (6-7.5V) applied across the device,  the
electron is accelerated towards the transparent contact while the
hole is accelerated towards the gain region.  The gain region is
moderately doped with As impurities, which are shallow impurities
lying only 54meV below the conduction band.  The device is cooled
to an operation temperature of 6-7K, so there is not enough
thermal energy to excite donor electrons into the conduction band.
These electrons are effectively frozen out in the impurity states.
However, when a hole is accelerated into the gain region it easily
impact ionizes these impurities, kicking the donor electrons into
the conduction band.   Scattered electrons can create subsequent
impact ionization events resulting in avalanche multiplication.

One of the nice properties of the VLPC is that, when an electron
is impact ionized from an As impurity, it leaves behind a hole in
the impurity state, rather than in the valence band as in the case
of APDs.  The As doping density in the gain region is carefully
selected such that there is partial overlap between the energy
states of adjacent impurities.  Thus, a hole trapped in an
impurity state can travel through conduction hopping, a mechanism
based on quantum mechanical tunnelling.  This conduction hopping
mechanism is slow, the hole never acquires sufficiency kinetic
energy to impact ionize subsequent As sites.  The only carrier
that can create additional impact ionization events is the
electron kicked into the conduction band.  Thus, the VLPC has a
natural mechanism for creating single carrier multiplication,
which is known to significantly reduce multiplication
noise~\cite{McIntyre66}.  We will return to this point in the
upcoming sections.

One of the disadvantages of using shallow As impurities for
avalanche gain is that these impurities can easily be excited by
room temperature thermal photons.  IR photons with wavelengths of
up to 30$\mu m$ can directly optically excite an impurity.  These
excitations can create extremely high dark count levels.  The
bi-layer structure of the VLPC helps to suppress this.  A visible
photon can be absorbed both in the intrinsic and doped silicon
regions.  An IR photon, on the other hand, can only be absorbed in
the doped region, as its energy is smaller than the bandgap of
intrinsic silicon.  Thus, the absorption length of IR photons is
much smaller than visible photons.  This suppresses the
sensitivity of the device to IR photons to about 2$\%$.  Despite
this suppression, the background thermal radiation is very bright,
requiring orders of magnitude of additional suppression.  In the
next section we will discuss how this is achieved.

\section{Cryogenic system for operating the VLPC}

  \begin{figure}
    \centerline{\scalebox{0.4}{\includegraphics{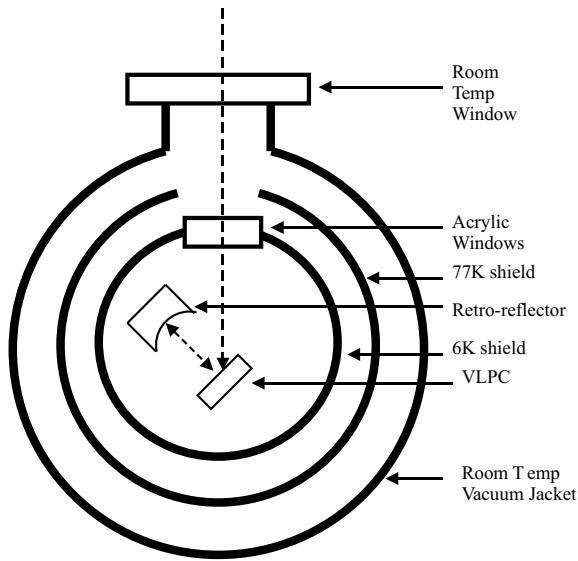}}}
    \caption{Schematic of cryogenic setup for VLPC.}\label{fig:VLPCcryo}
  \end{figure}

In order to operate the VLPC we must cool it down to cryogenic
temperatures to achieve carrier freezeout of the As impurities. We
must also shield it from the bright room temperature thermal
radiation which it is partially sensitive to.  This is achieved by
the cryogenic setup shown in Figure~\ref{fig:VLPCcryo}.

The VLPC is held in a helium bath cryostat.  As small helium flow
is produced from the helium bath to the cryostat cold finger by a
needle valve.  The helium bath is surrounded by a nitrogen jacket
for radiation shielding.  This improves the helium hold time.  A
thermal shroud, cooled to 77K by direct connection to the nitrogen
jacket, covers the VLPC and low temperature shielding.  This
shroud is intended to improve the temperature stability of the
detector by reducing the thermal radiation load.  A hole at the
front of the shroud allows photons to pass through.  The detector
itself is encased in a 6K shield made of copper.  The shield is
cooled by direct connection to the cold plate of the cryostat. The
front windows of the 6K radiation shield, which are also cooled
down to this temperature, are made of acrylic plastic. This
material is highly transparent at optical frequencies, but is
almost completely opaque from 2-30$\mu m$. The acrylic windows
provide us with the required filtering of room temperature IR
photons for operating the detector.  We achieve sufficient
extinction of the thermal background using 1.5-2 cm of acrylic
material.  In order to eliminate reflection losses from the window
surfaces, the windows are coated with a broadband anti-reflection
coating centered at 532nm.  Room temperature transmission
measurements indicate a 97.5$\%$ transmission efficiency through
the acrylic windows.  However, the performance of the
anti-reflection coating degrades when the windows are cooled down
to cryogenic temperatures.  Low temperature reflection
measurements indicate a 7$\%$ reflection loss.  This increased
loss is attributed to changes in the dielectric constant of the
material, resulting in a worse impedance match for the
anti-reflection coating.  Better engineering of the
anti-reflection coating could help eliminate these losses.

The surface of the VLPC has a broadband anti-reflection coating
centered around 550nm. Nevertheless, due to the large index
mismatch between silicon and air, there is still substantial
reflection losses on the order of $10\%$, even at the correct
wavelength. In order to eliminate this reflection loss, the
detector is rotated 45 degrees to the direction of the incoming
light.  A spherical refocussing mirror, with reflectance exceeding
99$\%$, is used to redirect reflected light back onto the detector
surface. A photon must reflect twice off of the surface in order
to be lost, reducing the reflections losses to less than 1$\%$.

The VLPC features high multiplication gains of about 30,000
electrons per photo-ionization event.  Nevertheless, this current
must be amplified significantly in order to achieve sufficiently
large signal for subsequent electronics.  The current is amplified
by a series of broadband RF amplifiers.  In order to minimize the
thermal noise contribution from the amplifiers, the first
amplification stage consists of a cryogenic pre-amplifier, which
is cooled to 4K by direct thermalization to the helium bath of the
cryostat.  The amplifier features a noise figure of 0.1 at the
operating frequencies of $30-500MHz$, with a gain of roughly 20dB.
The cryogenic amplifier is followed by additional commercial room
temperature RF amplifiers.  The noise properties of these
subsequent amplifiers is not as important since the signal to
noise ratio is dominated by the first cryogenic amplification
stage.  Using such a configuration, we achieve a 120mV pulse with
a 3ns duration when using 62dB of amplifier gain.

\section{Quantum efficiency and dark counts of the VLPC}

The quantum efficiency of the VLPC has already been studied at
650nm~\cite{TakeuchiKim99}.  Quantum efficiencies (QE) as high as
88$\%$ have been reported.  The dark counts at this peak QE were
20,000 1/s. Here we present measurements using a different
operating wavelength of 543nm, and a different cryogenic setup. As
a quick summary of what we will discuss, we observe raw quantum
efficiencies as high as 85$\%$ at this operating wavelength, with
dark count rates of roughly 20,000 1/s.  When correcting for
reflection losses from the windows and detector surface, we
estimate an intrinsic quantum efficiency of 93$\%$. These numbers
are consistent with previous measurements.

  \begin{figure}
    \centerline{\scalebox{0.4}{\includegraphics{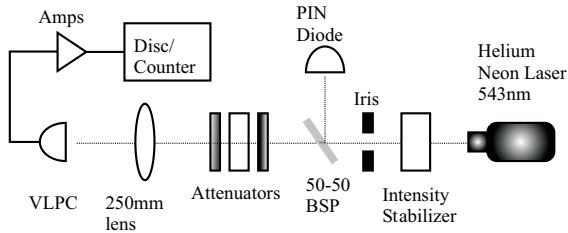}}}
    \caption{Experimental setup to measure quantum efficiency of the VLPC.}\label{fig:QEsetup}
  \end{figure}

The setup for measuring the quantum efficiency of the VLPC is
shown in Figure~\ref{fig:QEsetup}.  We use a helium neon laser
with an output wavelength of 543nm as a light source for the
measurement.  An intensity stabilizer is used to stabilize the
output of the laser to within about 0.1$\%$.  A 50-50 beamsplitter
is then used to send part of the laser to a calibrated PIN diode
to measure the power.  The power reading from the diode is
accurate to within a 2$\%$ calibration error.  Using this power
reading we can calculate the photon flux $N$, in units of photons
per second.  This is given by the relation
  \begin{equation}
    N = \frac{\lambda P}{hc},
  \end{equation}
where $\lambda$ is the wavelength of the laser, $P$ is the power
measured by the PIN diode, $h$ is Planke's constant, and $c$ is
the velocity of light in vacuum.

The laser is attenuated by a series of carefully calibrated
neutral density (ND) filters down to a flux of approximately
20,000 cps.  The attenuation required for this is on the order of
$10^{-9}$.  This flux is sufficiently small to ensure that we are
well within the linear regime of the VLPC.  At count rates
exceeding $10^{5}$ cps, the efficiency of the VLPC will begin to
drop due to dead time effects.  To measure the efficiency of the
VLPC we record the count rates of the detector, which we label
$N_c$, as well as the background counts $N_d$, which are measured
by blocking the laser.  The measured efficiency $\eta$ is given by
  \begin{equation}
    \eta = \frac{N_c-N_d}{\alpha N},
  \end{equation}
where alpha is the transmission efficiency of the ND filters.

  \begin{figure}
    \centerline{\scalebox{0.4}{\includegraphics{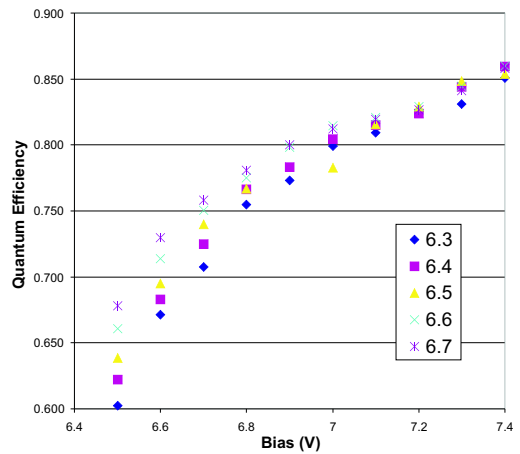}}}
    \caption{Quantum efficiency of VLPC vs. bias voltage for different
    temperatures.}\label{fig:QEvBias}
  \end{figure}

In Figure~\ref{fig:QEvBias}, we show the measured quantum
efficiency of the VLPC as a function of applied bias voltage
across the device.  Efficiencies are given for several different
operating temperatures.  At 7.4V bias the VLPC attains its highest
quantum efficiency of 85$\%$.  As the bias voltage is decreased
the quantum efficiency also decreases.  The reason for this is
that, at lower bias voltages, electrons created by impact
ionization of the initial hole are less likely to accumulate
sufficient kinetic energy in the gain region to trigger an
avalanche.  The bias voltage cannot be increased beyond 7.4V.
Beyond this bias the VLPC breaks down, resulting in large current
flow through the device.  This breakdown is attributed to direct
tunnelling of electrons from impurity sites into the conduction
band.

 \begin{figure}
    \centerline{\scalebox{0.4}{\includegraphics{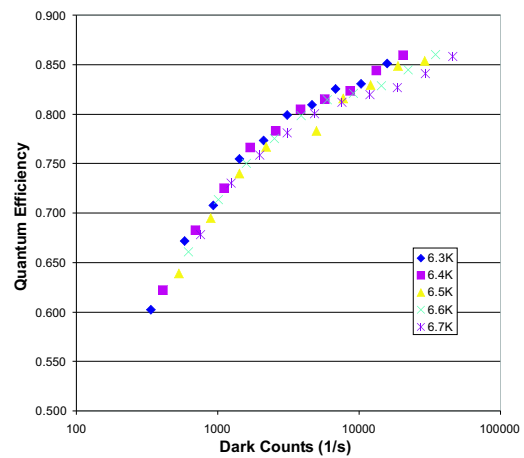}}}
    \caption{Quantum efficiency of VLPC vs. dark counts for different temperatures.}\label{fig:QEvDC}
  \end{figure}

One will notice that as the temperature is decreased, more bias
voltage is required to achieve the same quantum efficiency.  This
effect is attributed to a temperature dependance of the dielectric
constant of the device, which results in a change in the electric
field intensity in the gain region of the VLPC.  As the
temperature is decreased, it is speculated that the dielectric
constant increases, requiring higher bias voltage to achieve the
same electric field intensity.  This conjecture is supported by
the measurements shown in Figure~\ref{fig:QEvDC}.  In this figure
we plot the quantum efficiency as a function of dark counts,
instead of bias voltage.  Data is shown for the different
temperatures. Increasing the bias voltage results not only
increased quantum efficiency, but also in increased dark counts.
Increasing the temperature also increases both quantum efficiency
and dark counts.  But if we plot the quantum efficiency as a
function of dark counts, as is done in Figure~\ref{fig:QEvDC}, the
data for different temperatures all lie along the same curve. This
suggests that the quantum efficiency and dark counts both depend
on a single parameter, the electric field intensity in the gain
region. The temperature and bias voltage dependance of this
parameter result in the behavior shown in
Figure~\ref{fig:QEvBias}. From Figure~\ref{fig:QEvDC} we see that
the maximum quantum efficiency of 85$\%$ is achieved at a dark
count rate of roughly 20,000 cps.

In order to infer the actual efficiency of the VLPC alone, we must
correct for all other losses in our detection system.  The acrylic
windows are a big source of loss.  As mentioned previously, the
windows add a 7$\%$ reflection loss to our measurement. In
addition to this loss we have a reflection loss of 1$\%$ due to
the VLPC surface, despite the retro-reflector.  Other effects such
as detector dead time and beam focussing should contribute only
negligibly small corrections to the device efficiency.  Thus, the
efficiency of the VLPC detector itself is estimated to be $93\%$
at 543nm wavlengths.

\section{Noise properties of the VLPC}

When a photon is absorbed in a semi-conductor material, it creates
a single electron hole pair.  The current produced by this single
pair of carriers is, in almost all cases, too weak to observe due
to thermal noise in subsequent electronic components.  Single
photon counters get around this problem by using an internal gain
mechanism to multiply the initial pair into a much greater number
of carriers.  Avalanche photodiodes achieve this by an avalanche
breakdown mechanism in the depletion region of the diode.
Photomultipliers instead rely on successive scattering off of
dynodes.  The VLPC achieves this gain by impact ionization of
shallow arsenic impurities in silicon.

All of the above gain mechanism have an intrinsic noise process
associated with them.  That is, a single ionization event does not
produce a deterministic number of electrons.  The number of
electrons the device emits fluctuate from pulse to pulse.  This
internal noise is referred to as excess noise, or gain noise.  The
amount of excess noise that a device features strongly depends on
the mechanism in which gain is achieved.  The excess noise is
typically quantified by a parameter $F$, referred to as the excess
noise factor (ENF). The ENF is mathematically defined as
  \begin{equation}
    F = \frac{\langle M^2 \rangle}{\langle M \rangle^2},
  \end{equation}
where $M$ is the number of electrons produced by a
photo-ionization event, and the brackets notation represents a
statistical ensemble averages.  Noise free multiplication is
represented by $F=1$.  In this limit, a single photo-ionization
event creates a deterministic number of additional carriers.
Fluctuations in the gain process will result in an ENF exceeding
1.

The noise properties of an avalanche photo-diode are well
characterized.  The first theoretical study of such devices was
presented by McIntyre in 1966~\cite{McIntyre66}.  McIntyre studied
avalanche gain in the "Markov" limit.  In this limit, the impact
ionization probability for a carrier in the depletion region is a
function of the local electric field intensity at the location of
the carrier. In this sense, each impact ionization event is
independent of past history.  Under this assumption the ENF of an
APD was calculated. The ENF depends on the number of carriers that
can participate in the avalanche process.  If both electrons and
holes are equally likely to impact ionize, then $F\approx\langle M
\rangle$.  In the large gain limit the ENF is very big.
Restricting the impact ionization process to only electrons or
holes significantly reduces the gain noise.  In this ideal limit,
we have $F=2$.  This limit represents the best noise performance
achievable within the Markov approximation.

PMTs are known to have better noise characteristics than APDs. The
ENF of a typical PMT is around 1.2.  This suppressed noise is
because, in a PMT, a carrier is scattered off of a fixed number of
dynodes.  The only noise in the process is the number of electrons
emitted by each dynode per electron.

The multiplication noise properties of the VLPC have been
previously studied. Theoretical studies of the multiplication
noise have predicted that the VLPC should feature supressed
avalanche multiplication noise. This is due to two dominant
effects. First, because only electrons can cause impact
ionization, the VLPC features a natural single carrier
multiplication process. Second, the VLPC does not require high
electric field intensities to operate.  This is because impact
ionization events occur off of shallow arsenic impurities which
are only 54meV from the conduction band.  Thus, carriers do not
have to acquire a lot of kinetic energy in order to scatter the
impurity electrons. Because of the lower electric field
intensities, a carrier requires a fixed amount of time before it
can generate a second impact ionization.  This delay time
represents a deviation from the Markov approximation, and is
predicted to suppress the multiplication
noise~\cite{LaVioletteStapelbroek89}.  The ENF of the VLPC has
been experimentally measured to be less than 1.03
in~\cite{KimYamamoto97}.  Thus, the VLPC features nearly noise
free multiplication, as predicted by theory.  This low noise
property will play an important role in multi-photon detection,
which we discuss next.

\section{Multi-photon detection with the VLPC}

The nearly noise-free avalanche gain process of the VLPC opens up
the door to perform multi-photon detection.  When more than one
photon is detected by the VLPC, we expect the number of electrons
emitted by the detector to be twice that of a single photon
detection.  If the photons arrive within a time interval which is
much shorter than the electronic output pulse duration of the
detection system, then we expect to see a detection pulse which is
twice as high.

In the limit of noise free multiplication, this would certainly be
the case.  A single detection event would create $M$ electrons,
while a two photon event would create $2M$ electrons.  Higher
order photon number detections would follow the same pattern.
After amplification, the area or height of the detector pulse
would allow us to perfectly discriminate the number of detected
photons, even if they arrive on extremely short time scales.

In the presence of multiplication noise, the situation becomes
more complicated.  The pulse height of a one photon pulse will
fluctuate, as will that of a two photon pulse.  There becomes a
finite probability that we only detect one photon, but due to
multiplication noise the height of the pulse appears to be more
consistent with a two photon event, and vice versa.  Our ability
to discriminate the number of detected photons becomes a question
of signal to noise ratio.

There are ultimately two effects which will limit multi-photon
detection.  One is the quantum efficiency of the detector.  If we
label the quantum efficiency as $\eta$, then the probability of
detecting $n$ photons is given by $\eta^n$, assuming detector
saturation is negligible.  Thus, the detection probability is
exponentially small in $\eta$.  For larger $n$ this may produce
extremely low efficiencies.  The second limitation is the
electrical detection noise, as previously discussed.  There are
two contributions to the electrical noise.  One is the excess
noise of the detector, and the other is electrical noise
originating from amplifiers and subsequent electronics.  The
latter can in principle be eliminated by engineering ultra-low
noise circuitry.  The former, however, is a fundamental property
of the detector which cannot be circumvented, short of engineering
a different detector with better noise properties.

In the absence of detection inefficiency and amplifier noise, the
multiplication noise will ultimately put a limit on how many
simultaneous photons we can detect.  Defining $\sigma_m$ as the
standard deviation of the multiplication gain, the fluctuations of
an $n$ photon peak will be given by $\sqrt{n}\sigma_m$.  This is
because the $n$ photon pulse is simply the sum of $n$ independent
single photon pulses from different locations of the VLPC active
area.  Summing the pulses also causes the variance to sum,
resulting in the buildup of multiplication noise.  The mean pulse
height separation between the $n$ photon peak and the $n-1$ photon
peak, however, is constant.  It is simply proportional to $\langle
M \rangle$, the average multiplication gain.  At some sufficiently
high photon number, the fluctuations in emitted electrons will be
so large that there is little distinction between an $n$ and $n-1$
photon event.  We can arbitrarily establish a cutoff number at the
point where the fluctuations in emitted electrons is equal to the
average difference between an $n$ and $n-1$ photon detection
event.  In this limit, the maximum photon number we can detect is
  \begin{equation}
    N_{max} = \frac{1}{F-1}.
  \end{equation}
Using the above condition as a cutoff, we see that even an ideal
APD with $F=2$ cannot discriminate between 1 and 2 photon events.
A PMT with $F=1.2$ could potentially be useful for up to 5 photon
detection, but due to low quantum efficiencies of PMTS, this is
typically impractical.  The VLPC, with $F<1.03$ could potentially
discriminate more that 30 photons.  Furthermore it could
potentially do this with $93\%$ quantum efficiency.  However, this
limit is difficult to approach due to electronic noise
contribution from subsequent amplifiers.

\section{Characterizing multi-photon detection capability}

The multi-photon detection capability of the VLPC has been
previously studied.  Early studies used long light pulse
excitations, with poor electronic time resolution so that multiple
photons appeared as a single electronic pulse~\cite{AtacPark92}.
Later studies used twin photons generated from parametric
down-conversion, which arrive nearly simultaneous, to investigate
multi-photon detection~\cite{KimTakeuchi99}. These studies
restricted their attention to one and two photon detection. Higher
photon numbers were not considered.

  \begin{figure}
    \centerline{\scalebox{0.4}{\includegraphics{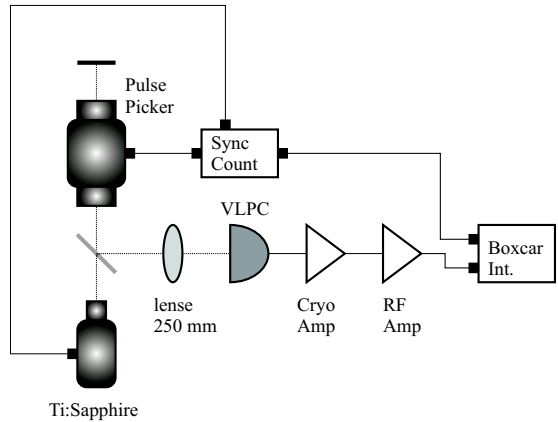}}}
    \caption{Experimental setup to measure multi-photon detection capability of
    the VLPC.}\label{fig:VLPCmultiSetup}
  \end{figure}

The experiment described below measures the photon number
detection capability of the VLPC when excited by multiple photons.
Figure~\ref{fig:VLPCmultiSetup} shows the experimental setup.  A
Ti:Sapphire laser, emitting pulses of about 3ps duration, is used.
The duration of the optical pulses are much shorter than the
electrical pulse of the VLPC detector, which is 2ns.  A pulse
picker is used to down-sample the repetition rate of the laser
from 76MHz to 15KHz. This is done in order to avoid saturation of
the detector.  A synchronous countdown module, which is used as
the pulse picking signal, is also used to trigger a boxcar
integrator.  The output of the VLPC is amplified by the amplifier
configuration discussed earlier.  The amplified signal is
integrated by a boxcar integrator.  The integrated value of a
pulse is proportional to the number of electrons emitted by the
detector, as long as amplifier saturation is negligible. The
output of the boxcar integrator is digitized by an analog to
digital converter, and stored on a computer.

 \begin{figure}
    \centerline{\scalebox{0.4}{\includegraphics{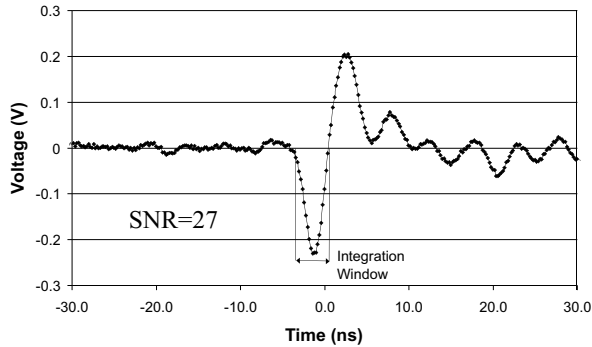}}}
    \caption{Oscilloscope pulse trace of VLPC output after room temperature
    RF amplifiers.}\label{fig:OscPulse}
  \end{figure}

Figure~\ref{fig:OscPulse} shows a sample oscilloscope pulse trace
of a VLPC pulse after the room temperature amplifiers.  The output
features an initial sharp negative peak of about 2ns full width at
the half maximum.  A positive overshoot follows.  This positive
overshoot is the result of the 30MHz high pass of the cryogenic
amplifiers.  If we compare the variance of the electrical
fluctuations before the pulse to the minimum pulse value, we
determine the signal to noise ration (SNR) to be 27.  The figure
also illustrates the integration window used by the boxcar
integrator, which captures only the negative lobe of the pulse.

\begin{figure*}
    \centerline{\scalebox{0.8}{\includegraphics{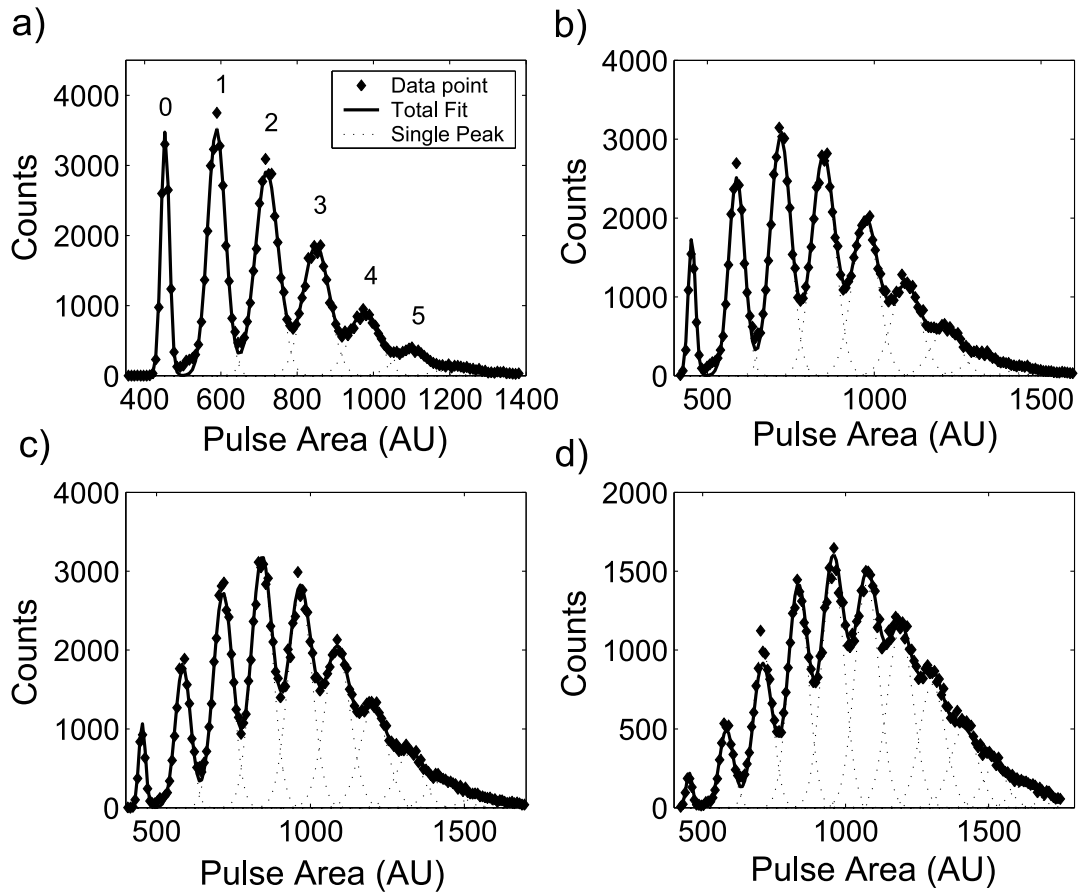}}}
    \caption{Pulse area spectrum generated by the boxcar
    integrator for four different excitation powers.  The dotted lines represent the fitted
    distribution of each photon number peak.  The solid line is the total sum of all the
    peaks.  Diamonds denote measured data points.  Each peak
    represents a photon number event, starting with zero photons
    for the first peak.}\label{fig:PAspectrum}
  \end{figure*}

In order to measure the multi-photon detection capability, we
attenuate the laser to about 1-5 detected photons per pulse. For
each laser pulse, the output of the VLPC is integrated and
digitized.  Figure~\ref{fig:PAspectrum} shows pulse area
histograms for four different excitation powers.  The area is
expressed in arbitrary units determined by the analog to digital
converter. Because the pulse area is proportional to the number of
electrons in the pulse, the pulse area histogram is proportional
to the probability distribution of the number of electrons emitted
by the VLPC.  This probability distribution features a series of
peaks. The first peak is a zero photon event, followed by one
photon, two photons, and so on.  In the absence of electronic
noise and multiplication noise, these peaks would be perfectly
sharp, and we would be able to unambiguously distinguish photon
number.  Due to electronic noise however, the peaks become
broadened and start to partially overlap.  The broadening of the
zero photon peak is due exclusively to electronic noise.  Note
that the boxcar integrator adds an arbitrary constant to the pulse
area, so that the zero photon peak is centered around 450 instead
of 0.  The one photon peak is broadened by both electronic noise
and multiplication noise. Thus, the variance of the one photon
peak is bigger than the zero photon peak. As the photon number
increases, the width of the pulses also increases due to buildup
of multiplication noise. This eventually causes the smearing out
of the probability distribution at around the seven photon peak.

In order to numerically analyze the results, we fit each peak to a
gaussian distribution.  Theoretical studies predict that the
distribution of the one photon peak is a bi-sigmoidal
distribution, rather than a
gaussian~\cite{LaVioletteStapelbroek89}. However, when the
multiplication gain as large, as in the case of the VLPC, this
distribution is well approximated by a gaussian. We use this
approximation because higher photon number events are sums of
multiple single photon events.  A gaussian distribution has the
nice property that a sum of gaussian distributions is also a
gaussian distribution.  In the limit of large photon numbers we
expect this approximation to get even better due to the central
limit theorem.

The most general fit would allow the area, mean, and variance of
each peak to be independently adjustable.  This allows too many
degrees of freedom, which often results in the optimization
algorithm falling into a local minimum.  To help avoid this, we do
not allow the average of each peak to be independently adjustable.
Instead, we require the averages to be equally spaced, as would be
expected from our model of the VLPC.  Thus, the average of the
i'th peak, denoted $x_i$, is determined by the relation
  \begin{equation}
    x_i = x_0 + i \Delta - i^2 \alpha.
  \end{equation}
In the above equation, $x_0$ is the average of the zero photon
peak, $\Delta$ is the spacing between peaks, and $\alpha$ is a
small correction factor which can account for effects such as
amplifier saturation.  These three parameters are all
independently adjustable. In all of our fits, $\alpha$ was much
smaller than $\Delta$ indicating the peaks are, for the most part,
equally spaced.

Figure~\ref{fig:PAspectrum} shows the results of the fits for each
excitation intensity.  The dotted lines plot the individual
gaussian distributions for the different photon numbers, and the
solid line plots the sum of all of the gaussians.  The diamond
markers represent the measured data points.
Table~\ref{table:FitResult} shows the center value and standard
deviation of the different peaks in panel c of the figure.  In
order to do photon number counting we must establish a decision
region for each photon number state.  This will depend, in
general, on the a-priori photon number distribution.  We consider
the case of equal a-priori probability, which is the worst case
scenario.  For this case, the optimal decision threshold between
two consecutive gaussian peaks is given by the point where they
intersect.  The value of this point can be easily solved, and is
given by,
  \begin{multline}
    x_d = x_i - \frac{\sigma_i^2 (x_{i+1} -
    x_i)}{\sigma_{i+1}^2-\sigma_i^2}\\
     + \frac{\sigma_i\sigma_{i+1}\sqrt{(x_{i+1} - x_i)^2- 2\left( \sigma_{i+1}^2-\sigma_i^2\right)
    \ln\frac{\sigma_i}{\sigma_{i+1}}}}{\sigma_{i+1}^2-\sigma_i^2}.
  \end{multline}
The probability of error for this decision is given by the area of
all other photon number peaks in the decision region.  This
probability is also shown in Table~\ref{table:FitResult}.

\begin{table}
\caption{Results of fit for panel (c) of
Figure~\ref{fig:PAspectrum}.} \label{table:FitResult}
  \begin{center}
  \begin{tabular}{c|c|c|c|c}
    Photon number & Avg. Area & Std. Dev. & $\%$Error  \\  \hline
    0 & 0 & 10.6 & 0.01  \\
    1 & 135 & 24.8 & 1.1 \\
    2 & 275 & 31.7 & 3.4 \\
    3 & 416 & 35.3 & 6.1 \\
    4 & 561 & 39.0 & 8.5 \\
    5 & 709 & 42.2 & 10.6 \\
    6 & 859 & 44.5 & 11.3
  \end{tabular}
  \end{center}
\end{table}

  \begin{figure*}
    \centerline{\scalebox{0.8}{\includegraphics{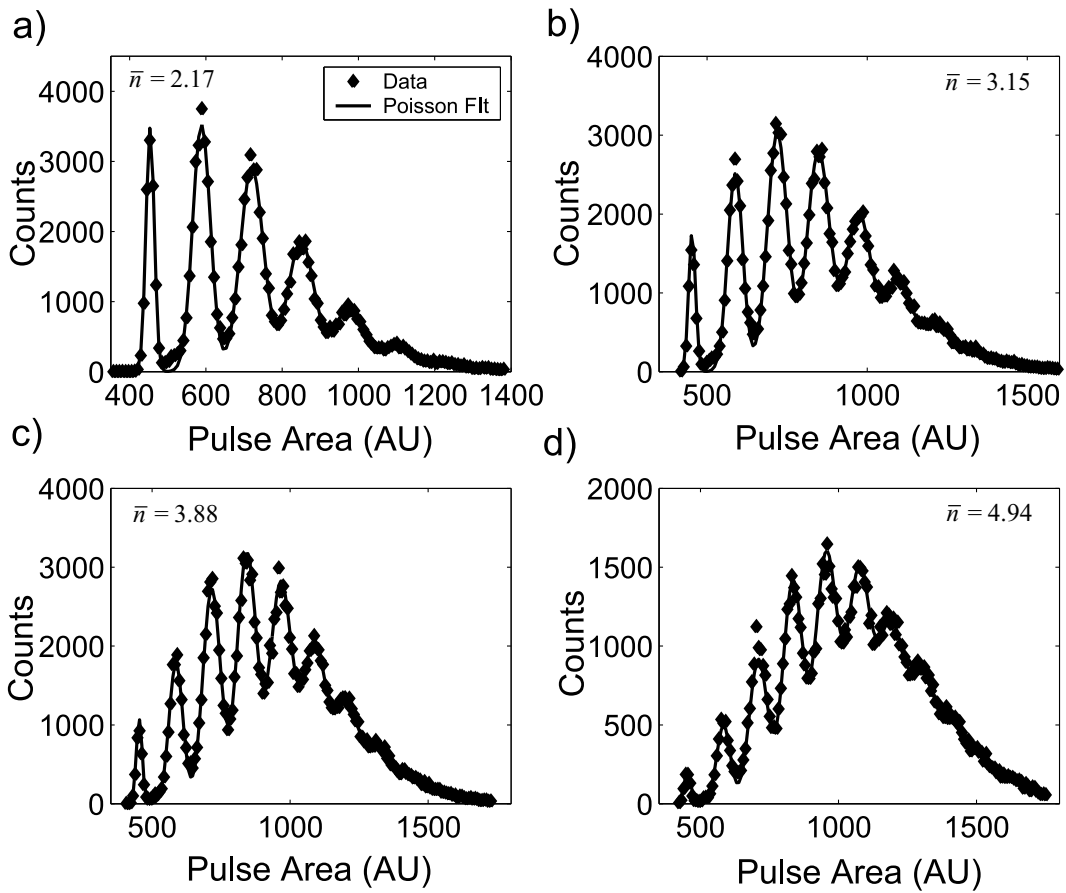}}}
    \caption{Pulse area spectrum fit to Poisson constraint on normalized
    peak areas.}\label{fig:PoissonSpec}
  \end{figure*}

From the data we would like to infer whether the VLPC is being
saturated at higher photon numbers.  If too many photons are
simultaneously incident on the detector, the detector surface may
become depleted of active area. This would result in a reduced
quantum efficiency for higher photon numbers.  In order to
investigate this possibility, we add an additional constraint to
the fit that the pulse areas must scale according to a Poisson
distribution.  Since the laser is a Poisson light source, we
expect this to be the case.  However, if saturation becomes a
factor, we would observe a number dependant loss.  This would
result in deviation from Poisson detection statistics.  In
Figure~\ref{fig:PoissonSpec} we plot the result of the fit when
the peak areas scale as a Poisson distribution.  One can see that
the imposition of Poisson statistics does not change the fitting
result in an appreciable way.  Thus, we infer that detector
saturation is not a strong effect at the excitation levels that we
are using.

\begin{figure}
    \centerline{\scalebox{0.4}{\includegraphics{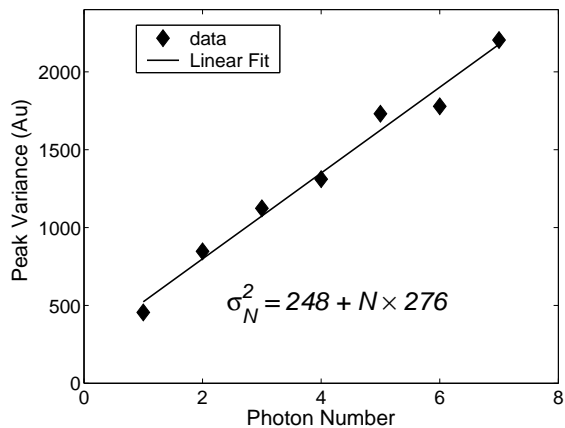}}}
    \caption{Variance as a function of photon number detection.  The
    linear relation is consistent with the independent detection
    model.}\label{fig:MultBuild}
  \end{figure}

The effect of multiplication noise buildup on the pulse height
spectrum can be investigated from the previous data.  In general,
we expect the pulse area variance to be a linearly increasing
function of photon number. This is consistent with the independent
detection model, in which an $n$ photon peak is a sum of $n$
single photon peaks coming from different areas of the detector.
To investigate the validity of this model, we plot variance as a
function of photon number in Figure~\ref{fig:MultBuild}. The
electrical noise variance, given by the zero photon peak, is
subtracted.  The variance is fit to a linear model given by
  \begin{equation}
    \sigma_i^2 = \sigma_0^2 + i \sigma_M^2.
  \end{equation}
In the above model, $i$ is the photon number, $\sigma_M^2$ is the
variance contribution from multiplication noise, and $\sigma_0^2$
is a potential additive noise term.  From the data, we obtain the
values $\sigma_M^2=276$, and $\sigma_0^2=246$.

A surprising aspect of this result is the large value of
$\sigma_0^2$. We expect that since electrical noise has been
subtracted, the only remaining contribution to the variance is
multiplication noise. If this were true, the value of $\sigma_0$
would be very small. Instead we obtain a value nearly equal to
that of $\sigma_m^2$. This may indicate that the electrical noise
is higher when the VLPC is firing, as opposed to when its not. A
change in the resistance of the device during the avalanche
process may effect the noise properties of subsequent
amplification circuits.  Further investigation is required in
order to determine whether this additive noise is fundamental to
the device, or can be eliminated in principle.

The above measurements of variance versus photon number gives us a
very accurate measurement of the excess noise factor $F$ of the
VLPC.  Previous measurements of $F$ for the VLPC have determined
that it is less than 1.03~\cite{KimYamamoto97}, which is nearly
noise free multiplication.  This number was obtained by measuring
the variance of the 1 photon peak, and comparing to the mean.
However, it is difficult to separate the electrical noise
contribution from the internal multiplication noise using this
technique.  Thus, the measurement ultimately determines only an
upper bound of $F$. By considering how the variance scales with
photon numbers, as we have done in Figure~\ref{fig:MultBuild}, the
multiplication noise can be accurately differentiated from
additive electrical noise. This allows us to calculate an exact
value for the excess noise factor.  From our measurement of
$\sigma_M^2$ and $\langle M \rangle$, we obtain an excess noise
factor of $F=1.015$.

\section{Conclusion}

In this paper we have discussed the interesting features of the
VLPC for quantum information processing.  The VLPC has the
potential to detect photons with quantum efficiencies approaching
93$\%$.  It also has the capability to do photon number detection,
a critical feature for linear optical quantum computation.  The
photon number detection capability of the VLPC is fundamentally
limited by internal noise processes in the device.  For many
applications, one does not require full photon number detection
capability.  It is sufficient to be able to distinguish between
one and more than one detection event.  The VLPC can do this with
99$\%$ accuracy.  Although the requirements for fully scalable
linear optical quantum computation are extremely demanding, the
VLPC may find use in areas where limited quantum computational
tasks are required.  Such fields as quantum cryptography and
quantum networking, where fully scalable computation is not always
required, may be able to incorporate the VLPC to perform novel
tasks.

\bibliographystyle{apsrev}
\bibliography{VLPC}

\end{document}